\documentclass[%
reprint,
nofootinbib,
 amsmath,amssymb,
 aps,
 pra,
]{revtex4-1}

\usepackage{graphicx}
\usepackage{dcolumn}
\usepackage{bm}
\usepackage{hyperref}

\usepackage[table]{xcolor}
\usepackage{ifthen}
\usepackage{mathrsfs}
\usepackage{amsmath} 
\usepackage{amssymb} 
\usepackage{amsthm}
\usepackage{booktabs}
\usepackage{url}
\usepackage{color}
\usepackage{multirow}
\usepackage{hyperref}
\definecolor{refcolor}{RGB}{0,15,255}
\hypersetup{
    colorlinks,
    citecolor=refcolor,
    filecolor=refcolor,
    linkcolor=refcolor,
    urlcolor=refcolor
}

\begin{document}
\newtheorem{theorem}{Theorem}[section]
\newtheorem{proposition}[theorem]{Proposition}
\newtheorem{lemma}{Lemma}[section]
\newtheorem{corollary}{Corollary}[section]
\newtheorem{definition}{Definition}[section]
\newtheorem{principle}{Principle}[section]
\newtheorem{hypothesis}{Hypothesis}[section]
\newtheorem{property}{Property}[section]
\newtheorem{problem}{Problem}
\newtheorem{objection}{Objection}
\theoremstyle{remark}
\newtheorem{remark}{Remark}[section]
\newtheorem{example}{Example}[section]
\newtheorem{reply}{Reply}

\newenvironment{solution}[2] {\paragraph{Solution to {#1} {#2} :}}{\hfill$\square$}
\newcolumntype{?}{!{\vrule width 1pt}}


\newcommand{\citepx}[1]{}

\newcommand{\todo}[1]{${\bf\Rightarrow}$ {\underline{\textbf{TODO}}} #1\PackageWarning{TODO:}{#1!}}
\def\({\left(}
\def\){\right)}
\def\[{\left[}
\def\]{\right]}

\newcommand{\tn}{\textnormal}
\newcommand{\ds}{\displaystyle}
\newcommand{\mf}[1]{\mathfrak{#1}}
\newcommand{\mc}[1]{\mathcal{#1}}
\newcommand{\ms}[1]{\mathscr{#1}}

\newcommand{\op}[1]{\operatorname{#1}}
\newcommand{\Tr}{\operatorname{Trace}}

\newcommand{\R}{\mathbb{R}}
\newcommand{\N}{\mathbb{N}}
\newcommand{\Z}{\mathbb{Z}}
\newcommand{\C}{\mathbb{C}}
\newcommand{\Oct}{\mathbb{O}}
\newcommand{\K}{\mathbb{K}}
\newcommand{\Hq}{\mathbb{H}}

\newcommand{\V}{\mathbb{V}}

\newcommand{\de}{\op{d}}
\newcommand{\diag}{\op{diag}}

\newcommand{\Id}{1}
\newcommand{\Null}{0}

\newcommand{\Dirac}{\op{D}_{\R}}
\newcommand{\DiracC}{\op{D}_{\C}}

\newcommand{\gammamod}{\Upsilon}
\newcommand{\tgamma}{\Gamma}

\newcommand{\tlambda}{\widetilde{\lambda}}

\newcommand{\Cl}{C\ell}
\newcommand{\CCl}{\C\ell}
\newcommand{\SMA}{\mc{A}_{\op{SM}}}
\newcommand{\SMAN}{\mc{A}_{\op{SM}0}^+}
\newcommand{\SMAAN}{\mc{A}_{\op{SM}1}^+}
\newcommand{\SMS}{\mc{S}_{\op{SM}}}
\newcommand{\D}{\mf{D}}

\newcommand{\mti}{m.t.i.}
\newcommand{\mtis}{m.t.i.s.}

\newcommand{\abs}[1]{|#1|}

\newcommand{\GO}{\op{O}}
\newcommand{\SO}{\op{SO}}
\newcommand{\so}{\mf{so}}
\newcommand{\GL}{\op{GL}}
\newcommand{\gl}{\mf{gl}}
\newcommand{\SL}{\op{SL}}
\newcommand{\sla}{\mf{sl}}
\newcommand{\U}{\op{U}}
\newcommand{\SU}{\op{SU}}
\newcommand{\uu}{\mf{u}}
\newcommand{\su}{\mf{su}}
\newcommand{\End}{\op{End}}
\newcommand{\Aut}{\op{Aut}}
\newcommand{\Spin}{\op{Spin}}
\newcommand{\spin}{\mf{spin}}
\newcommand{\Pin}{\op{Pin}}
\newcommand{\pin}{\mf{pin}}
\newcommand{\Span}{\op{span}}

\newcommand{\I}[1]{i_{\operatorname{#1}}}

\newcommand{\ChargeSpace}[1]{\chi_{\operatorname{#1}}}
\newcommand{\CSEM}{\ChargeSpace{em}}
\newcommand{\CSHyper}{\ChargeSpace{Y}}
\newcommand{\CSWeakIso}{\ChargeSpace{I_3}}
\newcommand{\CSColor}{\ChargeSpace{c}}
\newcommand{\CSElectroWeak}{\ChargeSpace{ew}}
\newcommand{\CSElectroColor}{\ChargeSpace{}}

\newcommand{\ChargeAntiSpace}[1]{\overline{\chi}_{\operatorname{#1}}}
\newcommand{\CSAEM}{\ChargeAntiSpace{em}}
\newcommand{\CSAHyper}{\ChargeAntiSpace{Y}}
\newcommand{\CSAWeakIso}{\ChargeAntiSpace{I_3}}
\newcommand{\CSAColor}{\ChargeAntiSpace{c}}
\newcommand{\CSAElectroWeak}{\ChargeAntiSpace{ew}}
\newcommand{\CSAElectroColor}{\ChargeAntiSpace{}}

\newcommand{\ssymEM}{\uu(1)_{\operatorname{em}}}
\newcommand{\ssymEMReal}{\so(2)_{\operatorname{em}}}
\newcommand{\ssymHyper}{\uu(1)_{\operatorname{Y}}}
\newcommand{\ssymWeak}{\su(2)_{\operatorname{L}}}
\newcommand{\ssymElectroWeakOld}{\symWeak\oplus\symHyper}
\newcommand{\ssymElectroWeak}{\uu(2)_{\operatorname{ew}}}
\newcommand{\ssymColor}{\su(3)_{\operatorname{c}}}

\newcommand{\SymEM}{\U(1)_{\operatorname{em}}}
\newcommand{\SymEMReal}{\SO(2)_{\operatorname{em}}}
\newcommand{\SymHyper}{\U(1)_{\operatorname{Y}}}
\newcommand{\SymWeakL}{\SU(2)_{\operatorname{L}}}
\newcommand{\SymWeak}{\SU(2)_{\operatorname{w}}}
\newcommand{\SymElectroWeakOld}{\SymWeak\times\SymHyper}
\newcommand{\SymElectroWeak}{\U(2)_{\operatorname{ew}}}
\newcommand{\SymColor}{\SU(3)_{\operatorname{c}}}

\newcommand{\ExtPow}[1]{\bigwedge{}^{#1}}
\newcommand{\ExtAlg}{\ExtPow{\bullet}}

\newcommand{\Matrix}[2]{{\mathbf{M}}_{#2}(#1)}

\newcommand{\tildeOp}{\mc{K}}
\newcommand{\JOp}{\mc{J}}

\newcommand{\clg}[1]{\gamma^{#1}}
\newcommand{\clq}[1]{\varsigma^{#1}}

\newcommand{\qfree}{\mf{q}}
\newcommand{\q}[1]{\qfree_{#1}}
\newcommand{\qd}[1]{\qfree{}\!^\dagger\!{}_{#1}}
\newcommand{\qc}[1]{\overline{\qfree}_{#1}}
\newcommand{\qcd}[1]{\qc{#1}^\dagger}
\newcommand{\qvol}{\qfree}
\newcommand{\qdvol}{\qvol{}\!^\dagger}
\newcommand{\h}{\mf{h}}
\newcommand{\cip}{\ms{C}}
\newcommand{\cim}{\widetilde{\cip}}
\newcommand{\w}[1]{\mf{u}_{#1}}
\newcommand{\wc}[1]{\overline{\w{}}_{#1}}

\newcommand{\qqd}{\wproj}
\newcommand{\qdq}{\wproj'}

\newcommand{\e}[1]{\mf{e}_{#1}}
\newcommand{\f}[1]{\tilde{\mf{e}}_{#1}}
\newcommand{\evol}{\e{}}
\newcommand{\fvol}{\f{}}

\newcommand{\bwfree}{\omega}
\newcommand{\bwindex}[1]{
{
  \ifthenelse{\equal{#1}{1}}{\operatorname{u}}{
  \ifthenelse{\equal{#1}{2}}{\operatorname{d}}{
  \ifthenelse{\equal{#1}{3}}{\circ}{#1}}}
}}
\newcommand{\bw}[1]{\bwfree_{\bwindex{#1}}}
\newcommand{\bwd}[1]{\bwfree{}\!^\dagger\!{}_{\bwindex{#1}}}
\newcommand{\bwvol}{\bwfree}
\newcommand{\bwdvol}{\bwvol{}\!^\dagger}

\newcommand{\we}[1]{\mf{u}_{\bwindex{#1}}}
\newcommand{\wf}[1]{\we{#1}'}
\newcommand{\wevol}{\we{}}
\newcommand{\wfvol}{\wf{}}

\newcommand{\wproj}{\mf{p}}

\newcommand{\wg}[1]{\widetilde{T}_{#1}}

\newcommand{\ns}{\ms{N}}
\newcommand{\nsd}{\ms{N}^\dagger}

\newcommand{\ws}[1]{\mathbb{W}_{#1}}
\newcommand{\wsu}{\ws{0}}
\newcommand{\wsd}{\ws{1}}
\newcommand{\wsuR}{\wsu{}_{\operatorname{R}}}
\newcommand{\wsuL}{\wsu{}_{\operatorname{L}}}
\newcommand{\wsdR}{\wsd{}_{\operatorname{R}}}
\newcommand{\wsdL}{\wsd{}_{\operatorname{L}}}
\newcommand{\wsw}{\ws{w}}
\newcommand{\wsn}{\ws{\circ}}
\newcommand{\WeakSpinor}[1]{\psi_{#1}}
\newcommand{\WeakUp}{\WeakSpinor0}
\newcommand{\WeakDown}{\WeakSpinor1}
\newcommand{\WeakUpL}{\WeakUp{}_{\operatorname{L}}}
\newcommand{\WeakUpR}{\WeakUp{}_{\operatorname{R}}}
\newcommand{\WeakDownL}{\WeakDown{}_{\operatorname{L}}}
\newcommand{\WeakDownR}{\WeakDown{}_{\operatorname{R}}}

\newcommand{\hns}{h_{\ns}}
\newcommand{\hw}{h_{\operatorname{w}}}

\newcommand{\ob}[1]{\mf{e}_{#1}}
\newcommand{\nbp}[1]{\mf{p}_{#1}}
\newcommand{\nbq}[1]{\mf{q}_{#1}}
\newcommand{\np}{\nbp{}}
\newcommand{\nq}{\nbq{}}
\newcommand{\ts}{\mc{V}}
\newcommand{\nsp}{\mc{P}}
\newcommand{\nsq}{\mc{Q}}
\newcommand{\spinorspace}[1]{W_{#1}{}}
\newcommand{\spinors}{\spinorspace{\np}}
\newcommand{\spinorsn}{\spinorspace{\overline{\nu}}}
\newcommand{\spinorsd}{\spinorspace{d}}
\newcommand{\spinorsu}{\spinorspace{\overline{u}}}
\newcommand{\spinorse}{\spinorspace{e}}

\newcommand{\vol}{\tn{vol}}

\newcommand{\qH}[1]{i_{#1}}
\newcommand{\qD}[1]{\mf{i}_{#1}}

\newcommand{\J}{\mf{j}}

\newcommand{\iMatrix}{\left(\begin{array}{rr}0 & -1 \\ 1 & 0\end{array}\right)}

\newcommand{\ie}{\textit{i.e.}}
\newcommand{\wrt}{\textit{w.r.t.}}
\newcommand{\cf}{\textit{cf.}}
\newcommand{\eg}{\textit{e.g.}}
\newcommand{\etc}{\textit{etc.}}

\newcommand{\sref}[1]{\S\ref{#1}}

\newcommand{\lleq}{\underline{\ll}}

\newcommand{\bra}[1]{\langle#1|}
\newcommand{\ket}[1]{|#1\rangle}
\newcommand{\braket}[2]{\langle#1|#2\rangle}
\newcommand{\Herm}{\textnormal{Herm}}

\newcommand{\hilbert}{\mathcal{H}}

\definecolor{Green}{RGB}{30,210,50}
\definecolor{Yellow}{RGB}{255,220,10}

\newcommand{\pcell}[5]{
\ifthenelse{\equal{#5}{+}}{
\ifthenelse{\equal{#1}{r}}{\cellcolor{red}}{\ifthenelse{\equal{#1}{y}}{\cellcolor{Yellow}}{\ifthenelse{\equal{#1}{b}}{\cellcolor{blue!80}}{\cellcolor{black!80}}}}{\ifthenelse{\equal{#2}{L}}{\color{white!100}{\mathbf{{#3}^{{}_{#1}}_{{}_{#2#4}}}}}{\color{white!100}{\mathbf{{#3}^{{}_{#1}}_{{}_{#2#4}}}}}}
}{
\ifthenelse{\equal{#1}{r}}{\cellcolor{Green}}{\ifthenelse{\equal{#1}{y}}{\cellcolor{violet!80}}{\ifthenelse{\equal{#1}{b}}{\cellcolor{orange}}{\cellcolor{gray!70}}}}{\ifthenelse{\equal{#2}{L}}{\color{white!100}{\mathbf{{\overline{#3}}^{{}_{\overline{#1}}}_{{}_{#2#4}}}}}{\color{white!100}{\mathbf{{\overline{#3}}^{{}_{\overline{#1}}}_{{}_{#2#4}}}}}}
}
}

\newcommand{\image}[3]{
\begin{figure}[ht]
\begin{center}
\includegraphics[width=#2\textwidth]{img/#1}
\caption{\small{\label{#1}#3}}
\end{center}
\end{figure}
}

\title{Chiral asymmetry in the weak interaction via Clifford Algebras}

\author{Ovidiu Cristinel Stoica}
\affiliation{
 Dept. of Theoretical Physics, NIPNE---HH, Bucharest, Romania. \\
	Email: \href{mailto:cristi.stoica@theory.nipne.ro}{cristi.stoica@theory.nipne.ro},  \href{mailto:holotronix@gmail.com}{holotronix@gmail.com}
	}%

\date{Feb 27, 2020}

\begin{abstract}
Despite its tremendous success, the Standard Model of particle physics does not explain why the weak interaction breaks chiral symmetry. Various unified theories got us closer to an answer, but too often the explanation consists of labeling the $\operatorname{SU}(2)_w$ singlet representations as right-handed, and the doublet ones as left handed. This by itself does not ensure a chiral preference, because chirality itself, arising  in the Dirac spinors, is not a property of the internal gauge group representations. Something deeper than mere labeling is required. In this paper, some of the progress using exterior and Clifford algebras is reviewed, and a possible explanation for chiral asymmetry is presented. It is shown how such a solution is present, rather implicitly, in a model proposed in a previous article.
\end{abstract}

\maketitle

\section{Introduction}
\label{s:intro}

An important question, unanswered in the Standard Model of particle physics, is \emph{Why is the electroweak interaction chiral?}

A general procedure to answer this question is to try to identify the rules behind the symmetry groups and their representations present in the Standard Model (SM), and then identify the mathematical structure behind these rules. This heuristic led to various models aiming to explain the properties of the SM from more fundamental structures.

In particular, a model that solves the problem of chirality of the electroweak interaction should solve at least three main problems.

The first problem is, of course,
\begin{problem}
\label{problem:weak_reps}
	The model must include the doublet and singlet representations of $\SymWeak$, as known from the Standard Model.
\end{problem}

The main problem to be solved to answer the question of the chiral preference of the electroweak interaction, is the following:
\begin{problem}
\label{problem:chiral}
The model must lead inevitably to the conclusion that leptons and quarks transform under $\SymWeak$ as doublets, in the left-handed case, and as singlets, in the right-handed case. Their anti-particles should transform under $\SymWeak$ as singlets, in the left-handed case, and as doublets, in the right-handed case.
\end{problem}

I used the notation $\SymWeak$ rather than the common one $\SymWeakL$, because the latter may ``smuggle'' into the model the idea that the weak interaction is automatically chiral.

While weak interaction seems to be easier to understand in the case of leptons, also quarks take part of it, so there is a third problem that should be solved along with these ones:
\begin{problem}
\label{problem:color_patterns}
The model must inevitably include the proper $\SymWeak$ as doublet or singlet representations for the up and down types of quarks, as well as for the charged and neutral leptons.
\end{problem}

Models addressing Problem \ref{problem:weak_reps} include \emph{Grand Unified Theories} (GUT) based on larger simple Lie groups like $\SU(5)$ \cite{georgi1974GUTSU5} and $\Spin(10)$ (sometimes referred to as $\SO(10)$) \cite{georgi1975GUTSO10,HaraldMinkowski1975GUTSO10}, models based on Clifford algebras \cite{Barducci1977QuantizedGrassmannVariablesGUT,Casalbuoni1979UnifiedQuarksLeptons,Casalbuoni1980unifiedQuarksLeptonsClifford,ChisholmFarwell1996CliffordAlgebrasForFundamentalParticles,Besprosvany2000GaugeSpaceTimeUnification,Daviau2015StandardModelCliffordAlgebra,Daviau2015RetourALOndeDeLouisDeBroglie,Daviau2015WeinbergAngle,Zenczykowski2013ElementaryParticlesEmergentPhaseSpace,Daviau2017StandardModelClifford15,Sto17StandardModelAlgebra,Zenczykowski2017CliffordAlgebraStandardModel,GillardGresnigt2019ThreeGenerationsCl8}, models based on octonions \cite{GunaydinGursey1974QuarkStatisticsOctonions,Dixon2013DivisionAlgebras,Furey2016StandardModelFromAlgebra}, sedenions \cite{GillardGresnigt2019ThreeGenerationsSedenions}, and the exceptional real Jordan algebra of dimension $27$ \cite{DuboisViolette2016JordanAlgebraGUT,TodorovDuboisViolette2017JordanAlgebraGUT}, without exhausting the list.
The models are different, but there are common structures appearing in different forms -- basically making use of the representations of $\SymWeak$ on the exterior algebra of a two-dimensional complex space, often this being expressed in terms of Clifford algebras.
Most proposed models simply combine the $\SymWeak$ doublet representation with the left-handed component of the Dirac spinor, and the $\SymWeak$ singlet representations with its right-handed component, and are seen sometimes as explaining chirality. A review of the common structure in terms of Clifford algebras, and of the general method for addressing Problem \ref{problem:weak_reps} can be found in \cite{Furey2018ElectroweakParity}, for the leptonic case.
However, getting the singlet and doublet $\SymWeak$ representations and labeling them as ``left-handed'' and ``right-handed'', and merely coupling these representations of the internal $\SymWeak$ with the correct chiral components of the Dirac spinor, is not enough to explain chiral asymmetry. Chirality is an observable of the Dirac spinors, not of the representations of $\SymWeak$, so a true explanation of the chiral preference of the electroweak interaction should also solve Problem \ref{problem:chiral}.

Some of the models mentioned above solve Problem \ref{problem:color_patterns} automatically, in particular the $\SU(5)$ and $\Spin(10)$ Grand Unified Theories\footnote{Although they require an additional symmetry breaking, additional interaction, and predict the proton decay, which was ruled out for reasonable parameters.},  the model in \cite{Casalbuoni1979UnifiedQuarksLeptons,Casalbuoni1980unifiedQuarksLeptonsClifford}, and \cite{Furey2016StandardModelFromAlgebra}.

An important step in answering Problem \ref{problem:chiral}, together with Problems \ref{problem:weak_reps}  and  \ref{problem:color_patterns}, is made in \cite{Casalbuoni1980unifiedQuarksLeptonsClifford}, in which it is shown how chiral spinors corresponding to both spacetime and internal degrees of freedom can be made to connect in the proper way the $\SymWeak$ representations with chirality.
A model addressing Problems \ref{problem:weak_reps} and \ref{problem:color_patterns}, and partially Problem \ref{problem:chiral}, can be found in \cite{Trayling1999GeometricApproachStandardModel,TraylingBaylis2001GeometricStandardModelGaugeGroup,TraylingBaylis2004Cl7StandardModel,Trayling2000AlgebraicStandardModelThesis}, although this model uses a representation of the Dirac spinors in which time is a scalar, and the chiral preference is obtained by imposing that right-handed neutrinos do not take part in any Standard Model interactions. A closely related model addressing Problems \ref{problem:weak_reps}, \ref{problem:chiral}, and \ref{problem:color_patterns}, and which at the same time includes Dirac spinors in the standard form, which is naturally Lorentz invariant, is proposed in \cite{Sto17StandardModelAlgebra}. There, the action of the Dirac algebra and of the spin group $\Spin_{1,3}$ break the symmetry, favoring a chiral component, and at the same time leads to a Weinberg angle given by $\sin^2\theta_W=0.25$. A central part of the present article consists of explaining the structures behind the model presented merely implicitly and in an undetailed form in \cite{Sto17StandardModelAlgebra}, and showing how this model address Problems \ref{problem:weak_reps}, \ref{problem:chiral}, and \ref{problem:color_patterns}.

Another problem, which will not be discussed here, is the existence of three generations, and some models deal with it, for example \cite{Casalbuoni1980unifiedQuarksLeptonsClifford,Furey2016StandardModelFromAlgebra}, and \cite{GillardGresnigt2019ThreeGenerationsSedenions,GillardGresnigt2019ThreeGenerationsCl8}, which extends \cite{Sto17StandardModelAlgebra} to three generations.

This article includes a discussion of the Clifford-algebra-based structures which allow to address the Problems \ref{problem:weak_reps}, \ref{problem:chiral}, and \ref{problem:color_patterns}, how various models do this partially, and a more explicit and detailed description of how this was implemented in \cite{Sto17StandardModelAlgebra} in an implicit way.
Section \sref{s:CliffordAlgPatternsSM} contains the relations between the representation of the Dirac algebra, and of the groups $\SymWeak$ and $\SymColor$, as they appear in the Standard Model, and the Clifford algebra representations. Section \sref{s:common_structure} explains how they connect to address the Problems \ref{problem:weak_reps}, \ref{problem:chiral}, and \ref{problem:color_patterns}. Section \sref{s:sma} describes how these problems are implicitly addressed by the model proposed in \cite{Sto17StandardModelAlgebra}. The article concludes with Section \sref{s:conclusions}.

\section{Clifford algebra patterns in the Standard Model}
\label{s:CliffordAlgPatternsSM}

This section explains some patterns in the Standard Model, which will be used to address Problems \ref{problem:weak_reps}, \ref{problem:chiral}, and \ref{problem:color_patterns}. We will see that the same underlying algebraic structure appears in three different forms, for each of the three problems.
In the case of Problems \ref{problem:weak_reps} and \ref{problem:color_patterns}, the structure arises from the  fundamental and the trivial representations of $\SymWeak$ and $\SymColor$, while in the case of Problem \ref{problem:chiral} from the representation of the Dirac algebra on one of its ideals. In a first form, the common structure is an exterior algebra, but more refinement leads to the idea that it is in fact a Clifford algebra. The fact that the same type of algebraic structure appears in all three instances, and the fact that these structures appear in one form or another in most proposed models, is not accidental.

\subsection{Exterior algebra patterns in the Standard Model}
\label{s:ExtAlgPatternsSM}

Both the weak symmetry group and the color symmetry group are instances of the special unitary group $\SU(n)$, for $n=2$, respectively $n=3$.
The standard representation of $\SU(n)$ is the complex vector space $\C_n\cong\C^n$ endowed with the canonical Hermitian metric. 
The fundamental and the trivial representations of $\SU(n)$ are $\ExtPow{k}\C_n$, and they are classified by charges of $\U(1)\subset\U(n)$. 
By examining the patterns of weak isospin and color of the leptons and quarks, we see that all of the fundamental and trivial representations of $\SymWeak$ and $\SymColor$ appear in the Standard Model.

The representations of the weak symmetry group $\SymWeak$ present in the Standard Model are displayed in Table \ref{tab:SymWeakSpaces}. They are all present, maybe with the exception of $\ExtPow{0}\C_2$, which corresponds to the still hypothetical right-handed neutrino.

\begin{table}[ht]
\begin{center}
\begin{tabular}{| c | c | r |}
\hline
Representation & Particles & Hypercharge\\
\hline
$\ExtPow{0}\C_2$ & $(\nu_{e})_R$ {\bf (?)} & $0$\\
$\ExtPow{1}\C_2$ & $(\nu_{e},e^-)_L$ & $-1$ \\
$\ExtPow{2}\C_2$ & $(e^-)_R$ & $-2$ \\
\hline
\end{tabular}
\caption{The representations of the weak symmetry group $\SymWeak$.}
\label{tab:SymWeakSpaces}
\end{center}
\end{table}

All of the fundamental and the trivial representations of the color symmetry group $\SymColor$ are present in the Standard Model, as seen in Table~\ref{tab:SymColorSpaces}.

\begin{table}[ht]
\begin{center}
\begin{tabular}{| c | c | r |}
\hline
Representation & Particles & Electric charge\\
\hline
$\ExtPow{0}\C_3$ & $\overline{\nu}_{e}$ & $0$\\
$\ExtPow{1}\C_3$ & $d^r,d^y,d^b$ & $-\frac 1 3$ \\
$\ExtPow{2}\C_3$ & $\overline u^{\overline{r}},\overline u^{\overline{y}},\overline u^{\overline{b}}$ & $-\frac 2 3$ \\
$\ExtPow{3}\C_3$ & $e^-$ & $-1$ \\
\hline
\end{tabular}
\caption{The representations of the color symmetry group $\SymColor$.}
\label{tab:SymColorSpaces}
\end{center}
\end{table}

An interesting difference between the representations listed in Table \ref{tab:SymWeakSpaces} and Table \ref{tab:SymColorSpaces} is that in the former, only particles appear, while the second one contains particles alternated with antiparticles. This remark is important to address Problem \ref{problem:color_patterns}.
In both cases, the conjugate representations are also valid, and they apply to the antiparticles of the particles from the tables.

\subsection{Dirac spinors and chirality}
\label{s:ExtAlgPatternsDirac}

Let $\gamma^\mu$ be the Clifford basis of the Dirac algebra $\CCl_{4}=\Cl_{1,3}\otimes\C$.

Following \cite{crumeyrolle1990clifford}, we define
\begin{equation}
\label{eq:dirac_witt_basis}
\begin{cases}
e_1 = \frac 1 2(\gamma_0+\gamma_3), e_2 = \frac 1 2(-i\gamma_2+\gamma_1)\\
f_1 = \frac 1 2(\gamma_0-\gamma_3), f_2 = \frac 1 2(-i\gamma_2-\gamma_1).\\
\end{cases}
\end{equation}

Then, $f_1f_2$ is nilpotent, and defines a minimal left ideal $\CCl_{4}f_1f_2$.

In the basis
\begin{equation}
\label{eq:dirac_weyl_basis}
(e_1 f_1f_2,e_2 f_1f_2,1 f_1f_2, e_1e_2 f_1f_2)
\end{equation}
of the minimal left ideal $\CCl_{4}f_1f_2$, the matrix form of $\gamma^\mu$ is the Weyl representation.

Let $\mc{W}$ be the vector space spanned by $(e_1,e_2)$.
Then, the spinors from $\ExtPow - \mc{W} f_1f_2$ are Weyl spinors of left chirality, and those from $\ExtPow + \mc{W} f_1f_2$ are Weyl spinors of right chirality.

Note that the basis \eqref{eq:dirac_weyl_basis} is in fact the same as the exterior algebra basis 
\begin{equation}
\label{eq:dirac_weyl_basis_ext}
(e_1 f_1 f_2,e_2 f_1f_2,1 f_1f_2, e_1 \wedge e_2 f_1f_2).
\end{equation}
This happens because the inner product between Dirac spinors vanishes on the space $\mc{W}$ spanned by $(e_1,e_2)$.
In fact, this is the reason why we can take as Dirac spinors the exterior algebra $\ExtAlg \mc{W}$, with the basis
\begin{equation}
\label{eq:dirac_weyl_basis_exterior}
(e_1,e_2,1, e_1 \wedge e_2).
\end{equation}
This fact will turn out to be relevant in the following.

\subsection{Representations of complex Clifford algebras, even dimension}
\label{s:cl_even}

We already noticed in subsection \sref{s:ExtAlgPatternsDirac} that there is an equivalence between the representation of the Dirac algebra $\CCl_{4}=\Cl_{1,3}\otimes\C$ by Clifford multiplication on one of its minimal left ideals, and a representation on the exterior algebra $\ExtAlg \mc{W}$.
This is in fact true for any complex Clifford algebra $\CCl_{2r}$ \cite{chevalley1997algebraicspinors,crumeyrolle1990clifford}.

To see this, let us consider the orthonormal basis of a complex $n$-dimensional vector space $V$ endowed with a complex inner product,
\begin{equation}
\label{eq:neutral_Clifford_basis}
(e_1,\ldots,e_r,e_{r+1},\ldots,e_{2r}),
\end{equation}
where $e_j^2=1$.

Then, we can build the \emph{Witt basis} basis 
\begin{equation}
\label{eq:complex_clifford_witt_basis_construction}
\begin{cases}
a_j := \frac 1 2(e_j+ie_{r+j}) \\
a_j^\dagger := \frac 1 2(e_j-ie_{r+j}). \\
\end{cases}
\end{equation}

The \emph{Witt decomposition} of the vector space $V$ is $V:=W\oplus W^\dagger$, where $W$ is spanned by $(a_j)$, and $W$ is spanned by $(a_j^\dagger)$.

Let $a\in\ExtPow r W$, $a:=a_1\wedge\ldots\wedge a_r=a_1\ldots a_r$.
Then, $a$ is \emph{nilpotent} ($a^2=0$), so $\ExtAlg W^\dagger a$ is a left ideal.
In fact, $\ExtAlg W^\dagger a$ is a minimal left ideal, and its elements are called \emph{algebraic spinors}.
It provides an irreducible representation of the Clifford algebra $\CCl_{2r}$, as we will see.

On the vector space $\ExtAlg W^\dagger a$ (hence also on the exterior algebra $\ExtAlg W^\dagger$), 
$a^\dagger$ and $a$ act like \emph{creation} and \emph{annihilation} operators:
\begin{equation}
\label{eq:complex_clifford_witt_basis}
\begin{cases}
\{a_j,a_k\} = \{a_j^\dagger,a_k^\dagger\} = 0 \\
\{a_j,a_k^\dagger\} = \delta_{jk}.
\end{cases}
\end{equation}

Let $\phi\in \ExtAlg W^\dagger$. Then,
\begin{equation}
\label{eq:complex_clifford_ladder}
\begin{cases}
a_j^\dagger \phi a = a_j^\dagger \wedge \phi a \\
a_j \phi a = i_{a_j}\phi a, \\
\end{cases}
\end{equation}
where $i_{a_j}$ is the \emph{interior product}.

Since $\dim W^\dagger=2^r$, this is an irreducible representation of $\CCl_{2r}$.

Because the inner product vanishes on $W$, the Clifford product on the subalgebra generated by $W$ coincides with the exterior product.
The same holds for $W^\dagger$.

The algebra $\CCl_{2r}$ is spanned by elements of the form
\begin{equation}
\label{eq:witt_clifford_basis}
a_{j_1}^\dagger \ldots a_{j_p}^\dagger a a^\dagger a_{k_1} \ldots a_{k_q},
\end{equation}
$p,q\in\{0,\ldots,r\}$, $1\leq j_1<\ldots<j_p\leq r$, $1\leq k_1<\ldots<k_q\leq r$ \cite{Sto17StandardModelAlgebra}.
Hence, the Clifford algebra $\CCl_{2r}$ is isomorphic to the matrix algebra $\C(2^r)$, according to the classification theory \cite{chevalley1997algebraicspinors,crumeyrolle1990clifford,law:1989}.

Since the elements of the form $a_{j_1}^\dagger \ldots a_{j_p}^\dagger$ span $\ExtAlg W^\dagger$,
$\CCl_{2r}$ is the direct sum of the minimal left ideals of the form 
\begin{equation}
\label{eq:complex_clifford_ideals}
\ExtAlg W^\dagger a a^\dagger a_{k_1} \ldots a_{k_q}.
\end{equation}

On these ideals, $\CCl_{2r}$ is represented just like on $\ExtAlg W^\dagger a$ \eqref{eq:complex_clifford_ladder}.

\subsection{From the exterior algebra \texorpdfstring{$\ExtAlg \C_r$}{} to Clifford algebra \texorpdfstring{$\CCl_{2r}$}{}}
\label{s:ext2cliff}

Conversely, we can start with the complex $r$-dimensional vector space $\C_r$ endowed with a Hermitian inner product $\mathfrak{h}$. Then, $\mathfrak{h}$ gives a canonical isomorphism $\overline{\C_r}\cong\C_r^\ast$.

On $\ExtAlg(\overline{\C_r}\oplus\C_r) \cong \ExtAlg \overline{\C_r}\otimes_\C\ExtAlg \C_r$, define an associative product by
\begin{equation}
\label{eq:clifford_from_exterior_and_inner_product}
\begin{cases}
uv :=u\wedge v, & u^\dagger v = u^\dagger \wedge v + \frac 1 2 u^\dagger(v), \\
u^\dagger v^\dagger :=u^\dagger \wedge v^\dagger, & u v^\dagger = u \wedge v^\dagger + \frac 1 2 v^\dagger(u),
\end{cases}
\end{equation}
for any $u,v\in\ExtAlg\C_r$.

Then, if $(a_j)$ is an orthonormal basis of $\C_r$,
\begin{equation}
\label{eq:complex_clifford_witt_basis_back}
\begin{cases}
\{a_j,a_k\} = \{a_j^\dagger,a_k^\dagger\} = 0 \\
\{a_j,a_k^\dagger\} = \delta_{jk}.
\end{cases}
\end{equation}

One obtains a Clifford algebra $\CCl(\overline{\C_r}\oplus\C_r)\cong\CCl_{2r}$.

The role of the Hermitian inner product $\mathfrak{h}$ consists in allowing the definition of the adjoint of an element $u\in\ExtAlg\C_r$.

Note that we don't need a Hermitian inner product $\mathfrak{h}$ if we start from $\ExtAlg(\C_r^\ast\oplus\C_r)$ rather than from $\ExtAlg(\overline{\C_r}\oplus\C_r)$, because we can define the Clifford product, instead of \eqref{eq:clifford_from_exterior_and_inner_product}, as
\begin{equation}
\label{eq:clifford_from_exterior_and_inner_product_no_Hermitian}
(u+\xi)(v+\eta) := (u+\xi)\wedge(v+\eta) + \frac 1 2 \(\xi(v) + \eta(u)\),
\end{equation}
for any $u,v\in\ExtAlg\C_r$ and $\xi,\eta\in\ExtAlg\C_r^\ast$.

\section{The common structure of the weak, color, and spin group representations}
\label{s:common_structure}

Let us make some remarks surrounding the discussion from Section \sref{s:CliffordAlgPatternsSM}, which will lead to a solution of Problems \ref{problem:weak_reps}, \ref{problem:chiral}, and \ref{problem:color_patterns}.

\begin{remark}
\label{rem:dirac_chiral}
The Dirac spinors are represented by $\ExtAlg\C_2$.
Left chiral spinors are represented by $\ExtPow{-}\C_2$, and right chiral spinors by $\ExtPow{+}\C_2$.
\end{remark}

\begin{remark}
\label{rem:ext_weak}
As seen in Table \ref{tab:SymWeakSpaces}, the weak force acts on the odd part $\ExtPow{1}\C_2=\ExtPow{-}\C_2$ of $\ExtAlg\C_2$ (which is not the same instance of $\ExtAlg\C_2$ as the one in Remark \ref{rem:dirac_chiral}).
A review of the occurrences of this pattern for the weak interaction in a few models, addressing Problem \ref{problem:weak_reps}, but not Problems \ref{problem:chiral} and \ref{problem:color_patterns}, was presented in \cite{Furey2018ElectroweakParity}
\end{remark}

\begin{remark}
\label{rem:dirac_weak}
Remarks \ref{rem:dirac_chiral} and \ref{rem:ext_weak} suggest that the exterior algebras used to represent Dirac spinors and the weak interaction are subalgebras of a larger space. On this space, the representation of $\SymWeak$ should leave unchanged the elements representing right chiral particles.
Since the representation of $\SymWeak$ on $\ExtAlg\C_2$ gives the correct doublet and singlets, to represent Dirac spinors we need to double the number of degrees of freedom. The representation of the Lorentz group, which is included in the Dirac algebra, should commute with the representation of $\SymWeak$.
\end{remark}

\begin{remark}
\label{rem:ext_color}
The representations of $\SymColor$ and $\SymEM$ on the exterior algebra $\ExtAlg\C_3$ and its conjugate are alternating particles with antiparticles (\emph{cf.} Table \ref{tab:SymColorSpaces}).
\end{remark}

\begin{remark}
\label{rem:ext_weak_color}
From Remarks \ref{rem:ext_weak} and \ref{rem:ext_color}, the representations of the Lie groups $\SymWeak$ and of $\SymColor$ should be combined into a larger space. The representations of $\SymWeak$ and $\SymColor$ on this larger space should commute. The natural combination is the exterior algebra $\ExtAlg \C_5$, which was used in \cite{georgi1974GUTSU5}. The groups $\SymWeak$ and $\SymColor$ were extended to the larger group $\SU(5)$, which is included in $\Spin(10)$. The group $\Spin(10)$ was used in \cite{georgi1975GUTSO10,HaraldMinkowski1975GUTSO10} with the same $\ExtAlg \C_5$ seen as a representation of $\Spin(10)$.
\end{remark}

\begin{remark}
\label{rem:dirac_ext_weak_color}
From Remarks \ref{rem:dirac_weak} and \ref{rem:ext_color}, the representations of the Dirac algebra, of $\SymWeak$, and of $\SymColor$, should be combined into a larger space. The representations of the Lorentz group, $\SymWeak$, and $\SymColor$ on this larger space should commute.
\end{remark}

\begin{remark}
\label{rem:dirac_ext_weak_color_spaces}
The space in Remark \ref{rem:dirac_weak} can be taken to be $\C_2\times\ExtAlg \C_2$, but also $\ExtAlg\C_1\times\ExtAlg \C_2$, which has the same dimension.
The space in Remark \ref{rem:dirac_ext_weak_color} can be taken to be $\ExtAlg\C_{1+2+3}$, in order to include all representations (which is larger than $\ExtAlg\C_{5}$ of the GUT models $\SU(5)$ and $\Spin(10)$). Note that, since the Dirac spinors include the antiparticles, it is enough to take for the colors either $\ExtAlg\C_3$, or its conjugate, and the same holds for $\ExtAlg\C_2$. The resulting dimension is the correct one for a generation of leptons and quarks, including their spin degrees of freedom.
\end{remark}

\begin{remark}
\label{rem:weak_cl6}
A way to obtain the extended representation from Remark \ref{rem:dirac_ext_weak_color_spaces} is to extend $\CCl_4$ to $\CCl_{6}$ (see Appendix B from \cite{Sto17StandardModelAlgebra}).
\end{remark}

\begin{remark}
\label{rem:gut_cl12}
One way to include the exterior algebra $\ExtAlg\C_6$ from Remark \ref{rem:dirac_ext_weak_color_spaces} is to extend the Dirac algebra from $\CCl_4$ to $\CCl_{12}$, and to use the representation of the spinors of $\CCl_{2r}$, where $r=6$, as in subsection \sref{s:cl_even}. This was proposed in \cite{Casalbuoni1980unifiedQuarksLeptonsClifford}. The representation on the space of spinors of $\CCl_{12}$ was also used later in \cite{Furey2016StandardModelFromAlgebra}, where a justification was given in terms of the Dixon algebra $\C\otimes\Hq\otimes\Oct$ (also see \cite{Dixon2013DivisionAlgebras}).
\end{remark}

\begin{remark}
\label{rem:gut_cl6}
There is a more compact way than the Clifford algebra $\CCl_{12}$ from Remark \ref{rem:gut_cl12}, which is still as natural.
The Clifford algebra $\CCl_{12}$ is represented as complex linear transformations of a $2^{\frac{12}{2}}$ complex vector space, the space of $\CCl_{12}$ spinors.
But, according to the algebraic definition of spinors, spinors are elements of a minimal left ideal of the Clifford algebra.
Since a Clifford algebra $\CCl_{2r}$ can be decomposed as a direct sum of $2^r$ minimal left ideals, {\cf} \eqref{eq:complex_clifford_ideals}, it follows that the total number of degrees of freedom representable as algebraic spinors is $2^r\times 2^r$. Hence, the number of degrees of freedom contained in the spinor of $\CCl_{12}$ is equal to the number of degrees of freedom contained in the sum of algebraic spinors \eqref{eq:complex_clifford_ideals} of the Clifford algebra $\CCl_6$.
\end{remark}

A model which represents the degrees of freedom of a generation of leptons and quarks was proposed in \cite{ChisholmFarwell1996CliffordAlgebrasForFundamentalParticles}. It includes the electroweak interactions based on $\Cl_{1,6}\cong\Cl_{1,3}\otimes\Cl_{0,3}$, and the ideals are interpreted as representing leptons and quarks, but it seems to lack explicit $\SymColor$ symmetry.
In \cite{Trayling1999GeometricApproachStandardModel,TraylingBaylis2001GeometricStandardModelGaugeGroup,TraylingBaylis2004Cl7StandardModel,Trayling2000AlgebraicStandardModelThesis}, the full symmetries of the Standard Model were shown to arise in a model based on $\Cl_7\cong\Cl_3\otimes\Cl_4$, from the condition to preserve the current and that the right-handed neutrino is non-interacting (which means that chiral asymmetry is imposed as a condition, but only for the neutrino, and it extends automatically to the electron and the quarks). The theory includes time as a scalar, and instead of four-vectors it employs \emph{paravectors} -- sums of three-vectors and scalars. While this allows to define Lorentz transformations, it does it by considering time as a scalar. In addition to the three space dimensions and the scalar time, which can be expressed using $\Cl_3$, there are four extra dimensions up to $\Cl_7$. The predicted Weinberg angle is given by $\sin^2\theta_W=0.375$.

A model which includes all leptons and quarks, together with their Dirac spin degrees of freedom, in the minimal left ideal decomposition of $\CCl_6$, was proposed in \cite{Sto17StandardModelAlgebra}. It shares some common structures, mentioned throughout this article, with other models, but it is different, and it was obtained independently and from different considerations. On the one hand, extending the Dirac algebra to include the weak symmetry $\SymWeak$ was shown in \cite{Sto17StandardModelAlgebra} to result in $\CCl_6$. The two actions require $\ExtAlg\C_3$, which can be obtained as a minimal left ideal of $\CCl_6$. On the other hand, starting from the remark that leptons and quarks, and their antiparticles, behave under $\SymColor$ like the exterior algebra $\ExtAlg\C_3$ and its conjugate (see Remark \ref{rem:ext_color} and Table \ref{tab:SymColorSpaces}), the tensor product of the two algebras $\ExtAlg\overline\C_3\otimes\ExtAlg\C_3$ turns out to be $\CCl_6$. This is because the algebra $\ExtAlg\C_3$ used to represent the color $\SymColor$ can be obtained as a minimal right ideal of the same Clifford algebra $\CCl_6$. The Dirac algebra and the weak $\SymWeak$ act at the left, and the color $\SymColor$ at the right, permuting the minimal left ideals. Since they act at the left and at the right, their actions commute. All minimal left ideals of $\CCl_6$ are then used to represent an entire generation of leptons and quarks. Then, the two ways to reach $\CCl_6$ correspond to actions on the left and on the right on the ideal decomposition of the algebra. The chiral preference of the weak interaction emerges due to the fact that the vector space $\overline\C_3\oplus\C_3$ has a natural orientation due to the Hermitian inner product $\mathfrak{h}$, combined with a geometric symmetry breaking introduced by the action of the Dirac algebra and of the spin group $\Spin_{1,3}$.
This resolves the necessity of introducing additional Higgs mechanisms of spontaneous symmetry breaking, and gives the correct number of degrees of freedom, and the correct symmetries, for a generation.

At the level of matrix representations of the algebras, the real Clifford algebras $\Cl_{1,6}$ and $\Cl_7$, and the complex Clifford algebra $\CCl_6$, are all represented by the same matrix algebra $\C(8)$, which explains the relation between the three models \cite{ChisholmFarwell1996CliffordAlgebrasForFundamentalParticles}, \cite{Trayling1999GeometricApproachStandardModel,TraylingBaylis2001GeometricStandardModelGaugeGroup,TraylingBaylis2004Cl7StandardModel,Trayling2000AlgebraicStandardModelThesis}, and \cite{Sto17StandardModelAlgebra}.
Note that, while the Clifford algebra $\CCl_6$ also occurs in octonion-based models like \cite{GunaydinGursey1974QuarkStatisticsOctonions,Dixon2013DivisionAlgebras} and \cite{Furey2016StandardModelFromAlgebra}, and its ideals are used to index leptons and quarks, and the $\SymColor$ and $U(1)$ symmetries result from $\CCl_6$, the minimal left ideals of this algebra are not used there to represent the full degrees of freedom of a generation of leptons and quarks. This makes sense indeed, since in these models octonions are fundamental and dictate the symmetries and the representations, and $\CCl_6$ occurs as operators acting on complex octonions. This is why the model from \cite{Furey2016StandardModelFromAlgebra} has to include the full algebra $\C\otimes\Hq\otimes\Oct$, and the spinor representation is in terms of $\CCl_{12}$, and not on the ideal decomposition of $\CCl_6$, despite its important role in the model for classifying the leptons and quarks.

In the following, I will show in more detail how the considerations made so far about the relation between chirality and the weak interaction are realized in the model \cite{Sto17StandardModelAlgebra}.

\section{The Standard Model Algebra}
\label{s:sma}

The construction from equation \eqref{eq:clifford_from_exterior_and_inner_product} was used in \cite{Sto17StandardModelAlgebra}, where we started from a complex three-dimensional vector space $\chi$ of the colors and obtained a representation of a generation of the Standard Model leptons and quarks, with the correct Dirac, Lorentz, electromagnetic, weak, and strong symmetries, with all required degrees of freedom, and the correct chiralities.

The representation of the color $\SymColor$ used in \cite{Sto17StandardModelAlgebra} corresponds to the representation in subsection \sref{s:ext2cliff} for $r=3$. This decomposes the Clifford algebra $\CCl_{6}$ into minimal left ideals as in equation \eqref{eq:complex_clifford_ideals}, of the form
\begin{equation}
\label{eq:complex_clifford_ideals_color}
\ExtAlg \chi^\dagger \qvol \qdvol\q{k_1} \ldots \q{k_j},
\end{equation}
where $(\q 1,\q 2,\q 3)$ is a basis of the vector space $\chi$, $j\in\{0,1,2,3\}$, $k_j\in\{1,2,3\}$, and
\begin{equation}
\label{eq:qvol}
\begin{cases}
\qvol :=& \q 1 \q 2 \q 3,\\
\qdvol :=& \qd 3 \qd 2 \qd 1, \\
\end{cases}
\end{equation}
which are nilpotent elements. The elements $\qqd:=\qvol\qdvol$ and $\qdq:=\qdvol\qvol$ are idempotent. Both the nilpotent and the idempotent elements defined here are primitive, and can be used to define minimal left or right ideals.

If, in \eqref{eq:complex_clifford_ideals_color}, $j=0$ or $j=3$, the minimal left ideals of the algebra represent pairs of charged and neutral leptons, and if $j=1$ or $j=2$ they represent up or down quarks, respectively their antiparticles. They transform like the correct singlet, respectively triplet representations of $\SymColor$ under the subgroup of the spin group of $\overline{\chi}\oplus\chi$ which preserves the Hermitian inner product on $\chi$ and $\chi^\dagger$. This result was based on the results from \cite{DoranHestenes1993LGasSG}. The same was obtained in \cite{Furey2016StandardModelFromAlgebra}, building on the work of \cite{GunaydinGursey1974QuarkStatisticsOctonions,Barducci1977QuantizedGrassmannVariablesGUT,Casalbuoni1979UnifiedQuarksLeptons}, and it is based on preserving the Witt decomposition, which in our case is $\overline{\chi}\oplus\chi$. Both representations of $\SymColor$ are equivalent and include, in addition to the color symmetry, the $\SymEM$ electromagnetic symmetry.

Since the right action of $\SymColor$ is different than the left actions of the Dirac algebra and of $\SymWeak$, it is justified to choose, for the minimal left ideal $\ExtAlg \chi^\dagger q$, a different basis than the one based on $q_{j}$ used to index the minimal left ideals in \eqref{eq:complex_clifford_ideals_color}.
The purpose is to make explicit the left actions of the Dirac algebra and of $\SymWeak$.

For this, let us define first
\begin{equation}
\label{eq:basis_sma}
\begin{cases}
\e j = \q j + \qd j \\
\f j = i\(\qd j - \q j\), \\
\end{cases}
\end{equation}
where $j\in\{1,2,3\}$, and also define
\begin{equation}
\label{eq:efvol}
\begin{cases}
\evol :=& \e 1 \e 2 \e 3, \\
\fvol :=& \f 1 \f 2 \f 3. \\
\end{cases}
\end{equation}

We now define
\begin{equation}
\label{eq:wewg-vs-ef}
	\left\{\begin{array}{llll}
		\we 1 &=-i\e3\f1, &  \wf 1 &=\fvol\e2, \\
		\we 2 &=i\fvol, &  \wf 2 &=i\evol, \\
		\we 3 &=i\f2\e3, &  \wf 3 &=\fvol\e1, \\
	\end{array}\right.
\end{equation}
where $j\in\{\bwindex1,\bwindex2,\bwindex3\}$, and
\begin{equation}
\label{eq:basis_w_back}
\begin{cases}
\bw j = \frac12\(\we j + i\wf j\) \\
\bwd j = \frac12\(\we j - i\wf j\), \\
\end{cases}
\end{equation}

The representations of $\SymWeak$ from Table \ref{tab:SymWeakSpaces} are then spanned by the various products involving $\bwd 1$ and $\bwd 2$, which are the singlets spanned by $1$ and $\bwd 1 \wedge \bwd 2 = \bwd 1 \bwd 2$, and the doublet spanned by $\bwd 1$ and $\bwd 2$. But since to each particle also correspond two degrees of freedom coming from the dimension of the Weyl spinor, one needs a way to double the degrees of freedom, according to Remark \ref{rem:dirac_weak}, and this is provided by $1$ and $\bwd 3$. From Remark \ref{rem:dirac_ext_weak_color_spaces} we learn that the inclusion of both leptons and quarks requires the exterior algebra $\ExtAlg\C_3$, which occurs as $\ExtAlg \chi$, and multiplies the minimal left ideal at the right.

We consider the minimal left ideal determined by the idempotent element $\wproj$ as split into two four-dimensional space of Dirac spinors,
\begin{equation}
\begin{cases}
\label{eq:spinor_space_dirac}
\wsu:=\Span_\C\(\wproj,\bwd3\wproj,\bwd1\wproj,\bwd1\bwd3\wproj\)\\
\wsd:=\Span_\C\(\bwd2\wproj,\bwd3\bwd2\wproj,\bwd1\bwd2\wproj,\bwd1\bwd3\bwd2\wproj\).
\end{cases}
\end{equation}

We also define the following subspace of Weyl spinors,
\begin{equation}
\label{eq:spinor_space_weyl}
\ws{w}:=\Span_\C\(\wproj,\bwd3\wproj\)
\end{equation}

Then, a generation of leptons and quarks is classified as in Table \ref{tab:standard_model_fermions_sma}, as shown in \cite{Sto17StandardModelAlgebra}. 

By inspecting Table \ref{tab:standard_model_fermions_sma}, we can see that the degrees of the bases from the algebras $\ExtAlg\C_{2}$ and $\ExtAlg\C_{3}$ used to represent the symmetries $\SymWeak$ and $\SymColor$ implement the Remarks from Section \sref{s:common_structure}, by this addressing the Problems \ref{problem:weak_reps}, \ref{problem:chiral}, and \ref{problem:color_patterns}.
Full details showing how they transform under the $\SymColor$, $\SymWeak$, and $\SymEM$ groups are given in \cite{Sto17StandardModelAlgebra}, which addresses implicitly the Problems \ref{problem:weak_reps}, \ref{problem:chiral}, and \ref{problem:color_patterns}. Here we explained the underlying mechanisms leading to the fact that the weak force is chiral as we know it for both leptons and quarks, as well as their antiparticles, which was rather implicit in \cite{Sto17StandardModelAlgebra}.

\section{Conclusions}
\label{s:conclusions}

We have seen the underlying mechanisms connecting chirality with the weak force and the notion of particle-antiparticle. We have seen how these relations are accounted for by representations on exterior algebras and Clifford algebras. The chiral preference of the weak interaction does not simply come from calling the singlet representations right-handed and the doublet ones left-handed, or from merely adding the subscript $L$ to $\SU(2)$, or even simply declaring that only the left chiral part of the Dirac spinor has to participate in the weak interaction. This points out to the idea that there should be an intimate relation between the Dirac spinors and the internal degrees of freedom, geometric in nature. We have seen that to have an answer, one has to address Problems \ref{problem:weak_reps}, \ref{problem:chiral}, and \ref{problem:color_patterns}, and how these problems were addressed to various degrees in several models. In all these examples, the exterior algebra or the Clifford algebra representations turned out to be essential. An account of how the weak force turns out to be chiral, which is implicit and automatically present in the model proposed in \cite{Sto17StandardModelAlgebra} was given more explicitly here.

\onecolumngrid

\def\arraystretch{1.25}
\begin{table}[ht]
\begin{center}
\resizebox{\columnwidth}{!}{%
    \begin{tabular}{ ? l l ? r | r | r | r ? r | r | r | r ? p{3cm} |}
    \bottomrule[1.5pt]
    \textbf{Particle} & & $\mathbf{\nu}$ & $\mathbf{\overline d}$ & $\mathbf{u}$ & $\mathbf{e^+}$ & $\mathbf{e^-}$ & $\mathbf{\overline u}$ & $\mathbf{d}$ & $\mathbf{\overline \nu}$ \\\hline
    \textbf{Spinor space} & & $\wsu$ & $\wsu\q{j}$ & $\wsu\q{jk}$ & $\wsu\q{123}$ & $\wsd$ & $\wsd\q{j}$ & $\wsd\q{jk}$ & $\wsd\q{123}$ \\ \hline
    \textbf{Classifier} & & $\wproj$ & $\wproj\q{j}$ & $\wproj\q{jk}$ & $\wproj\q{123}$ & $\qdvol\wproj$ & $\qdvol\wproj\q{j}$ & $\qdvol\wproj\q{jk}$ & $\qdvol\wproj\q{123}$ \\ \hline
    \textbf{Electric charge} & & $0$ & $+\frac{1}{3}$ & $+\frac{2}{3}$ & $+1$ & $-1$ & $-\frac{2}{3}$ & $-\frac{1}{3}$ & $0$ \\ \hline
		\multirow{2}{*}{\textbf{Chiral space}}
		& \textbf{L} & $\bwd1\ws{w}$ & $\ws{w}\q{j}$ & $\bwd1\ws{w}\q{jk}$ & $\ws{w}\q{123}$ & $\bwd2\ws{w}$ & $\bwd1\bwd2\ws{w}\q{j}$ & $\bwd2\ws{w}\q{jk}$ & $\bwd1\bwd2\ws{w}\q{123}$ \\
		& \textbf{R} & $\ws{w}$ & $\bwd1\ws{w}\q{j}$ & $\ws{w}\q{jk}$ & $\bwd1\ws{w}\q{123}$  & $\bwd1\bwd2\ws{w}$ & $\bwd2\ws{w}\q{j}$ & $\bwd1\bwd2\ws{w}\q{jk}$ & $\bwd2\ws{w}\q{123}$ \\ \hline
		\multirow{2}{*}{\textbf{Weak isospin}}
		& \textbf{L} & $+\frac{1}{2}$ & $0$ & $+\frac{1}{2}$ & $0$ & $-\frac{1}{2}$ & $0$ & $-\frac{1}{2}$ & $0$ \\
		& \textbf{R} & $0$ & $+\frac{1}{2}$ & $0$ & $+\frac{1}{2}$ & $0$ & $-\frac{1}{2}$ & $0$ & $-\frac{1}{2}$ \\ \hline
		\multirow{2}{*}{\textbf{Hypercharge}}
		& \textbf{L} & $-1$ & $+\frac{2}{3}$ & $+\frac{1}{3}$ & $+2$ & $-1$ & $-\frac{4}{3}$ & $+\frac{1}{3}$ & $0$ \\
		& \textbf{R} & $0$ & $-\frac{1}{3}$ & $+\frac{4}{3}$ & $+1$ & $-2$ & $-\frac{1}{3}$ & $-\frac{2}{3}$ & $1$ \\ \hline
		\toprule[1.25pt]
    \end{tabular}
}
\end{center}
\caption{Properties of leptons and quarks in the model based on the Clifford algebra $\CCl_6$ from \cite{Sto17StandardModelAlgebra}.}
\label{tab:standard_model_fermions_sma}
\end{table}
\twocolumngrid

\providecommand{\bysame}{\leavevmode\hbox to3em{\hrulefill}\thinspace}
\providecommand{\MR}{\relax\ifhmode\unskip\space\fi MR }
\providecommand{\MRhref}[2]{%
  \href{http://www.ams.org/mathscinet-getitem?mr=#1}{#2}
}
\providecommand{\href}[2]{#2}

\end{document}